\documentclass[twocolumn,showpacs]{revtex4}

\usepackage{amsmath, amssymb}
\usepackage{dcolumn}
\usepackage{latexsym}

\usepackage{indentfirst}
\usepackage{amssymb}
\usepackage{tabularx}
\usepackage{supertabular}
\usepackage{graphicx}

\begin{document}

\title{Relation between dry granular flow regimes and morphology of deposits: formation of lev\'ees in pyroclastic deposits}

\author{Gwena\"{e}lle F\'elix}
\author{Nathalie Thomas}
\altaffiliation[Corresponding author: ]{IUSTI-CNRS, 5 rue E. Fermi, Technop\^ole de Ch\^ateau-Gombert, 13453 Marseille, France. Tel: +33 4 91 10 68 72; Fax: +33 4 91 10 69 69. {\it E-mail:} nathalie.thomas@polytech.univ-mrs.fr}
\affiliation{Laboratoire Magmas et Volcans, CNRS,\\
5 rue Kessler, 63 000 Clermont-Ferrand, France}

\begin{abstract}
Experiments on dry granular matter flowing down an inclined plane are performed in order to study the dynamics of dense pyroclastic flows. The plane is rough, and always wider than the flow, focusing this study on the case of laterally unconfined (free boundary) flows.
We found that several flow regimes exist depending on the input flux
and on the inclination of the plane. Each flow regime corresponds to a particular morphology of the associated deposit. In one of these regimes, the flow reaches a steady state, and the deposit exhibits a lev\'ee/channel morphology similar to those observed on small pyroclastic flow deposits. The lev\'ees result from the combination between lateral 
static zones on each border of the flow and the drainage of 
the central part of the flow after the supply stops. Particle segregation features
are created during the flow, corresponding to those observed on the deposits of pyroclastic flows. 
Moreover, the measurements of the deposit morphology (thickness of the channel, height of the lev\'ees, width of the deposit) are quantitatively related 
to parameters of the dynamics  of the flow (flux, velocity, height of the flow), leading to a way of studying the flow dynamics from only measurements of the deposit. Some attempts to make extensions to natural cases are discussed through experiments introducing the polydispersity of the particle sizes and the particle segregation process.

\vspace{3mm}

\end{abstract}


\maketitle

\section{Introduction}

Flows of grains are encountered in many natural situations: sand dunes, debris flows, landslides, turbidity currents, avalanches... These flows involve complex interactions between grains, and sometimes a fluid. 
For example, the dynamics of pyroclastic flows can follow every possible combination between two end-members, depending on relative intensities of particle-particle, and gas-particle interactions:  dilute suspensions (only gas-particle interactions), and granular flows (only particle-particle interactions). If large volume pyroclastic flows are related to gravity currents (dense or dilute) \cite{fisher, dade, Branney}, small volume flows behave as granular flows, produced either by dome or by eruptive column collapse \cite{Calder, Nairn, Ui, Yamamoto, colima, Cole2002}.
Many models have been proposed (reviews in \cite{freundt, legros}), ranging from dilute turbulent suspension flows dominated by fluid-particle interactions \cite{fisher, dade, bursik, Woods94, Sparks93, huppert}, to dense gravity currents (concentrated suspensions) with fluidization effects and some particle-particle interactions \cite{sparks, wilsonmodel, roche, burgisser, Ishida, takahashi}, and to flows controlled by particle collisions \cite{takahashi}. But, to our knowledge, there is no study involving frictional interactions between particles.

In this paper, we focus on dense dry granular flows. They are defined as a flowing layer of grains, in which air-grain interactions are negligible (consequently, heat exchanges do not influence processes). For dense granular flows, interactions between grains are dominated by friction \cite{GDR, olivierchevoir} and not by collisions, as it is the case for rapid granular flows \cite{Campbell90, drake, Azanza}. Moreover, if the flow is said ``dry" all cohesive forces between the grains are assumed negligible. Their dynamics is different from those of gravity currents which obey fluid mechanics, and in which grains are, at least partly, suspended by the vertical component of the fluid velocity. Gravity currents loose particles by sedimentation, and get dilute, unlike dense granular flows whose volume fraction of grains remains close to 60\% \cite{GDR}. Some pyroclastic flows have intermediate dynamics (a granular layer flow below a gravity current), and pyroclastic granular flows are surrounded by a suspension of ash. The aim of this paper is to obtain laws for the dynamics of the `dense granular flows' end-member, and quantitative relations between their dynamics and the morphological characteristics of their deposits. In that way, it should be possible to get quantitative information about flow velocities from pyroclastic deposit measurements.

\begin{figure}[htbp]
\begin{minipage}[t]{7.5cm}
\center
\includegraphics[width=7.5cm]{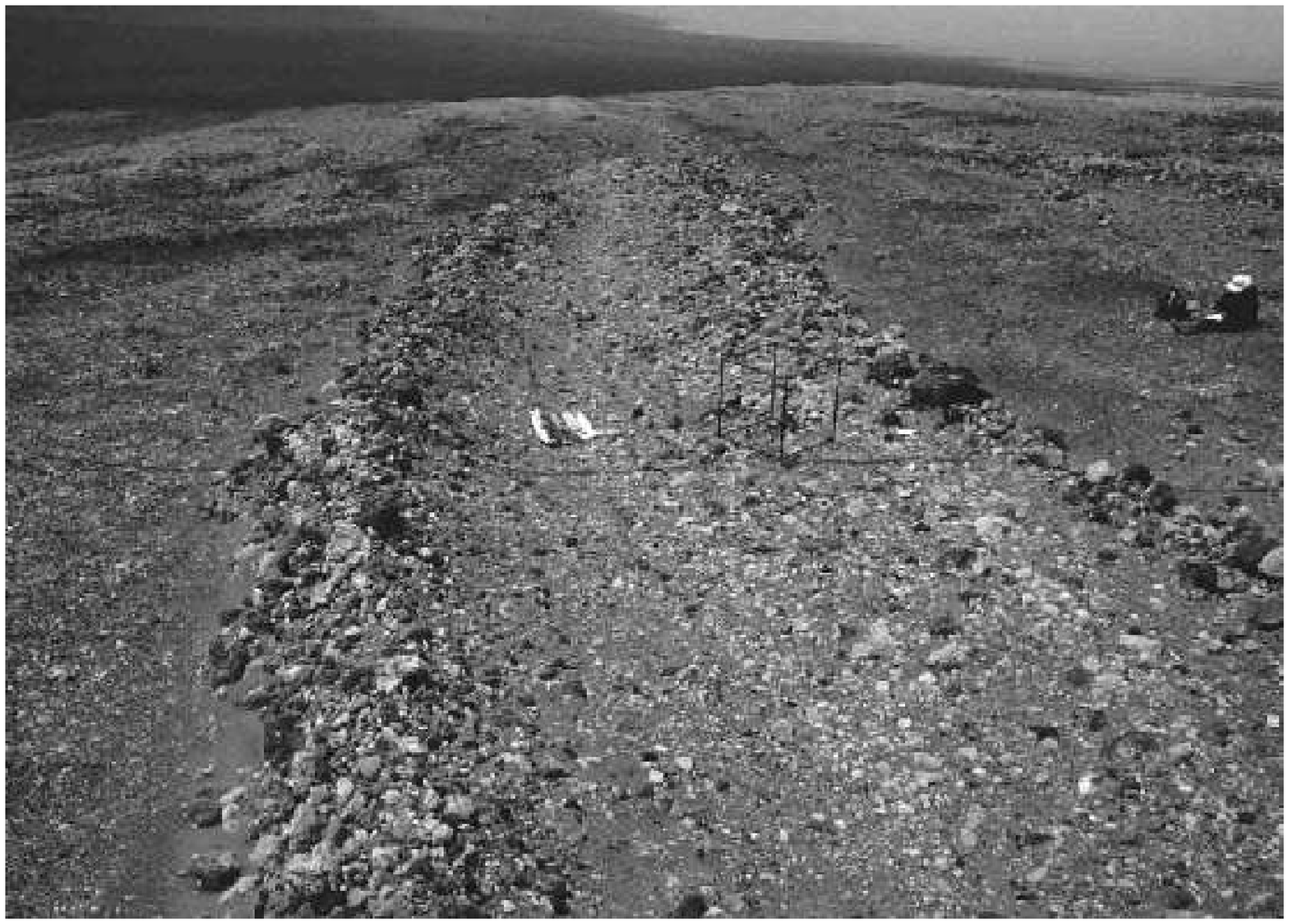} 
\end{minipage}
\hfill
\begin{minipage}[t]{7.5cm}
\includegraphics[width=7.5cm]{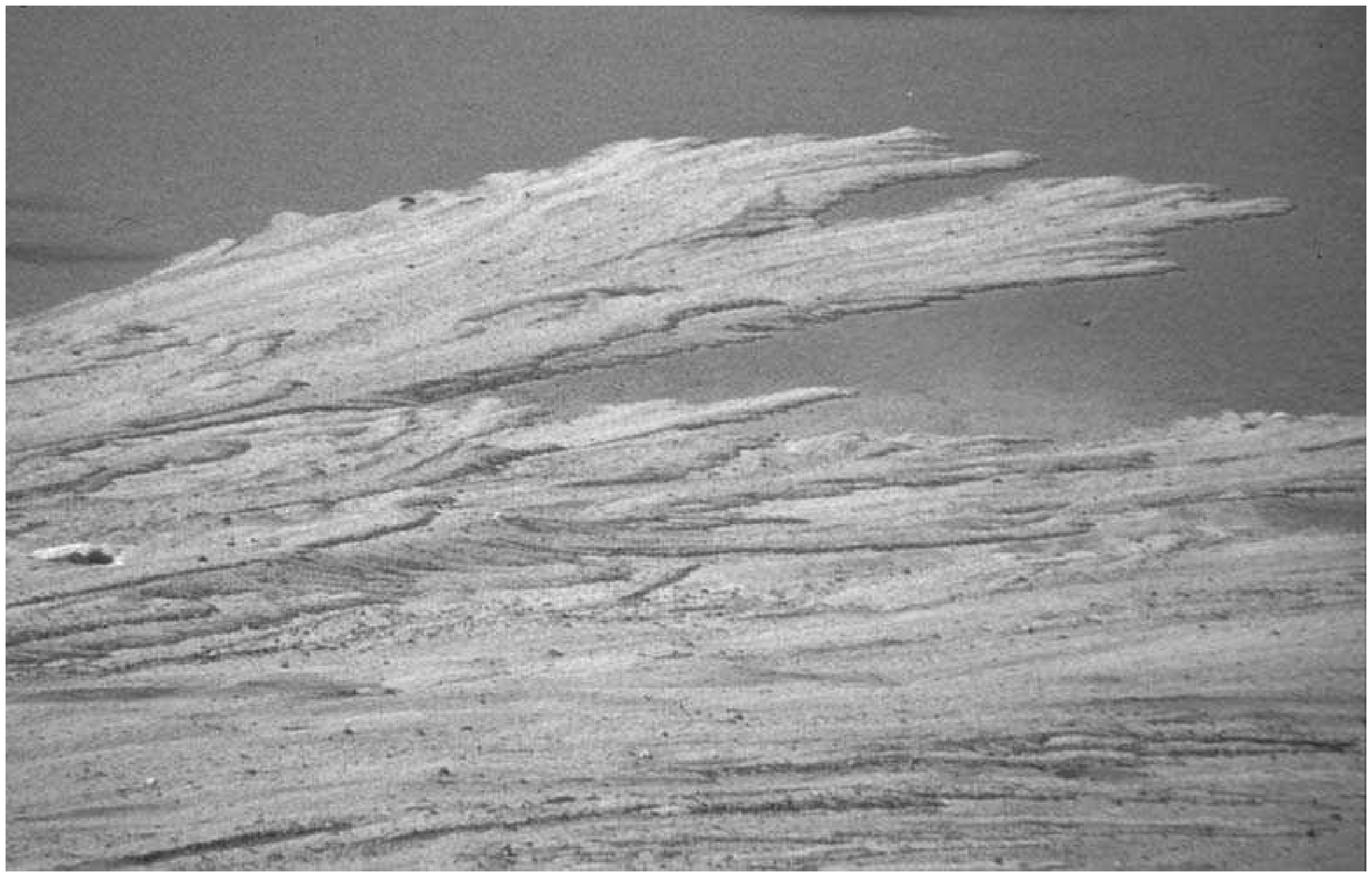}
\end{minipage}
\caption{Lobe-shaped deposits produced by the 1993 pyroclastic flows of Lascar volcano (Chile):  (top) lev\'ees enriched in large pumices (flow moving ahead), and (bottom) global view of lobes observed on the South-East flank of the volcano. The cube gives a 3D scale: edges are 78 cm long (photo from authors).}
\label{Lascar}
\end{figure}

The assumption about the occurence of dense pyroclastic granular flows, or their dense granular basal part, is mainly supported by deposit observations: segregation features (fig. \ref{Lascar}), erosion due to the friction between clasts \cite{Grunewald}, and erosion of the subtratum due to the dense character of the flow \cite{Calder, erosionlascar}. Moreover, there are no pyroclastic deposits on high slopes although they appear when slope decreases \cite{davies, Yamamoto, Calder}. When not laterally confined in valleys, the deposits display a particular morphology: the lobes have high parallel lateral lev\'ees (about 1 m high) enriched in large blocks while the central channel is lower and mainly composed of smaller particles \cite{davies, Rowley, wilson, Yamamoto, Miller, Ui, Calder, colima, Cole2002, cole98, Rodriguez, Beget89, Nairn}. The fronts have rounded bulbous shapes and steep margins (fig. \ref{Lascar}) \cite{Nairn, Rowley, erosionlascar}. Lobes deposited on the lowest slopes exhibit similar shapes. Various hypothesis have been proposed to explain the lev\'ee/channel morphology: 
lateral static zones linked to a Bingham rheology \cite{wilson, davies, Yamamoto}, drainage 
of the central part of the deposit with static lev\'ees \cite{Rowley}, differential deflation after emplacement \cite{Rowley} and differential fluidization of 
borders during emplacement \cite{wilson}, both due to the heterogeneity in the spatial distribution of particle sizes. This includes interaction of the borders with ambiant air creating differential flotation and segregation \cite{wilson}, and rafts of pumice clasts stranded on the sides of the flow \cite{Calder}. One aim of this paper is to know which part of these morphology and segregation features is due to the dense flow of dry granular matter.

Most experimental studies of dry granular flows concern confined flows; these flows are either confined in narrow channels \cite{savagesegregation, Azanza, ancey, jean, bonamy}, or in a wide geometry set-up where friction at lateral walls is negligible \cite{olivier}. The width of the flow is imposed by the width of the experimental set-up: the flow and deposit morphologies are strongly affected by confinement (in a similar way, except a few studies on snow avalanches \cite{Beghin}, experiments on gravity currents have been done in channels, in which the transverse morphology of the deposit is impossible to study).
Some studies focus on the spreading of a finite mass of granular matter on a wide slope \cite{Savage, oliviercalotte}, but no steady state regime is reached, and the deposit morphology is strongly influenced by its initial shape, or deformed by local changes of the slope \cite{gray}. Experiments involving the destabilisation of a bump of material, irregularly fed during about a quarter of the flow duration, show that the lev\'ee/channel morphology can develop \cite{McDonald}. But, to our knowledge there is no data on dynamics of unconfined dry granular flows on inclines, that are fed with a constant flux for a long duration. Models based on kinetic theory \cite{drake, Azanza, Savage} are valid for rapid collisional flows, although depth-averaging equations with a basal friction seem more appropriate for dense frictional flows \cite{olivier} (review in \cite{olivierchevoir}). All these studies show that there is no layer with a zero vertical gradient of the horizontal velocity at the surface of the flow, eliminating a Bingham rheology.

In this study, laboratory experiments involving dense granular flows have been performed in order to relate possible flow regimes and associated deposit morphologies. The input flux is constant such that the flow reaches a steady state. The flow is never as wide as the set-up, and its width is free to adapt to the dynamical conditions. Precise experiments have been carried out to understand the mechanism of formation of the lev\'ee/channel morphology and to quantitatively link information on the flow dynamics and measurements of the deposit morphology.

\section{Experimental apparatus}

\begin{figure}[htbp]
\center
\includegraphics[width=7.5cm]{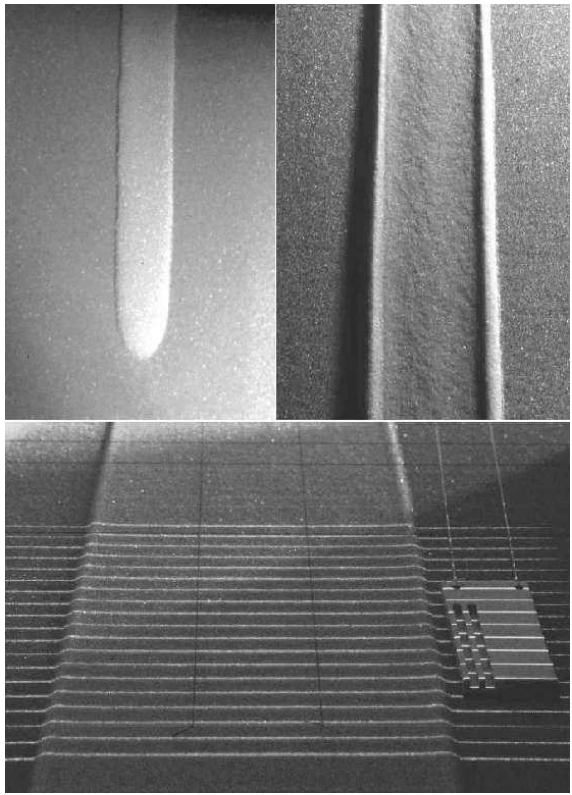}
\caption{(top, left) A glass beads finger-shaped flow on a rough 80 cm wide and 2 m long inclined plane, and (top, right) the associated deposit lighted from the right, showing lateral lev\'ees and low central channel. The width is around 12 cm in each case. (bottom) The deviation of the laser sheets, projected perpendicularly to the flow direction, is proportional to the deposit and flow thicknesses. The scale is given by the calibrated holes on the small plate (right of the picture); the width of this deposit is around 30 cm.}
\label{photolevee}   
\end{figure}

The flows are created by release of microbeads on an inclined plane (fig. \ref{photolevee}). The plane (80 cm wide, 2 m long) is made rough by gluing 425-600 $\mu$m beads on it. The slope is chosen such that the flow is dominated by frictional interactions between the beads, and between the beads and the plane, and not by collisions \cite{Campbell90, yoelrouleaux}. Two feeding systems are used: (1) A funnel with a variable aperture allows control of the flux imposed by a point source producing unconfined finger-shaped flows. (2) A box with a 40 cm gate is used to create sheet flows, rapidly becoming laterally confined, i.e. 80 cm wide flows.

Particles are 2.5 g/cm$^3$ density glass beads. Their diameter is either in a narrow (but not monodisperse), or in a wider (named `polydisperse' here) range of sizes. The humidity in the laboratory is held between 50\% and 55\% to avoid electrostatic effects or capillary bridges between particles, which are cohesive effects \cite{Fraysse}. 

The morphologies of flows and associated deposits are measured by projecting several laser sheets on the plane, perpendicularly to the flow direction (fig. \ref{photolevee}, bottom). The deviation of each line is proportional to the thickness of granular  material on the plane. A reference allows the scaling of the transverse profiles. 

For each set of experimental conditions, the width $L$ and the thickness $h$ of the flow as well as the deposit morphology are measured. The propagation of the flow is followed in time, giving the front velocity $v$, that corresponds to the mean flow velocity.

			\section{The lev\'ee/channel morphology and others} 

First, the mode of formation of lev\'ees in granular flow deposits is presented. Then, the observed flow regimes are described and related to the morphology of the associated deposits. Finally, we show how it is possible to get quantitative information on the flow dynamics, using only measurements of the deposit morphology.

			\subsection{Formation of lev\'ees and channel}\label{levee}

\begin{figure}[htbp]
\center
\includegraphics[width=7.5cm]{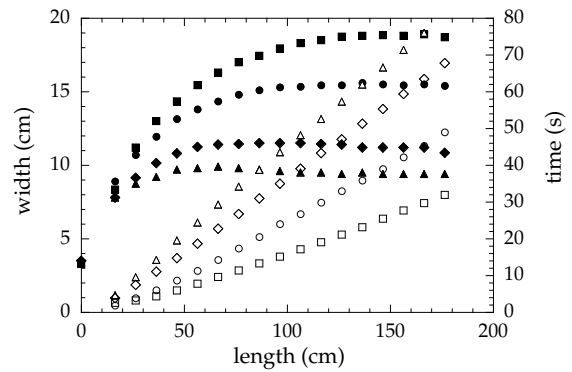}
\caption{For fluxes of: ({\large$\blacktriangle$}, $\triangle$) 6 g/s, ($\blacklozenge$, $\lozenge$) 8.5 g/s, ({\Large$\bullet$}, {\Large$\circ$}) 
19 g/s, and ({\footnotesize{$\blacksquare$, $\square$}}) 34 g/s: (plain symbols) longitudinal variation of the width of the flow down the slope (length is the distance from the source). The flow reaches its stable configuration at a distance between 50 and 150 cm from the source depending on the flux; (open symbols) variation of the position of the front of the flow (i.e. length of the flow) versus time. The linear evolution, after a first slightly curved part, shows that the velocity of the flow is constant once it has reached its stable width (beads: 350-390 $\mu$m, roughness: 425-600 $\mu$m, slope: 25$^\circ$).}
\label{largeurdist}
\end{figure}

Particles constituting the flows are in a narrow range of sizes (typically 300-400 $\mu$m). The flow develops from the point source. When moving downwards, it spreads laterally until it reaches a constant width. Once its stable width reached, the position of the front varies linearly with time: the velocity is constant (fig. \ref{largeurdist}). The flux being constant, the thickness of flowing material is also constant and the flow reaches a steady state regime (partial derivatives of velocity, width, height, with respect to time are equal to 0). 
When fully developed, the flow is finger-shaped with parallel borders and a curved bulbous front (fig. \ref{photolevee}, top, left), corresponding to the shape of deposits in the field \cite{wilson}.

\begin{figure}[htbp]
\includegraphics[width=7.5cm]{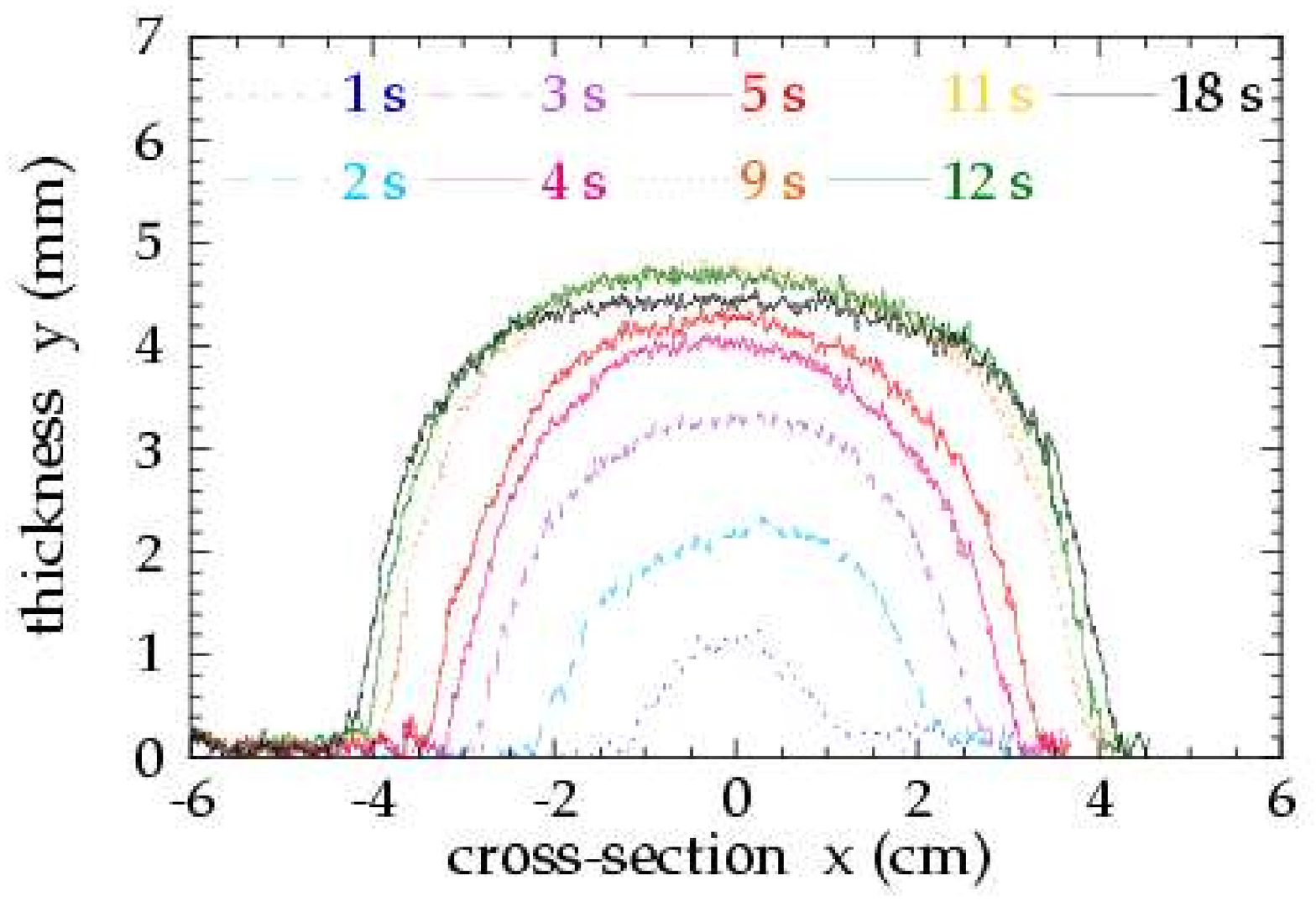}
\includegraphics[width=7.5cm]{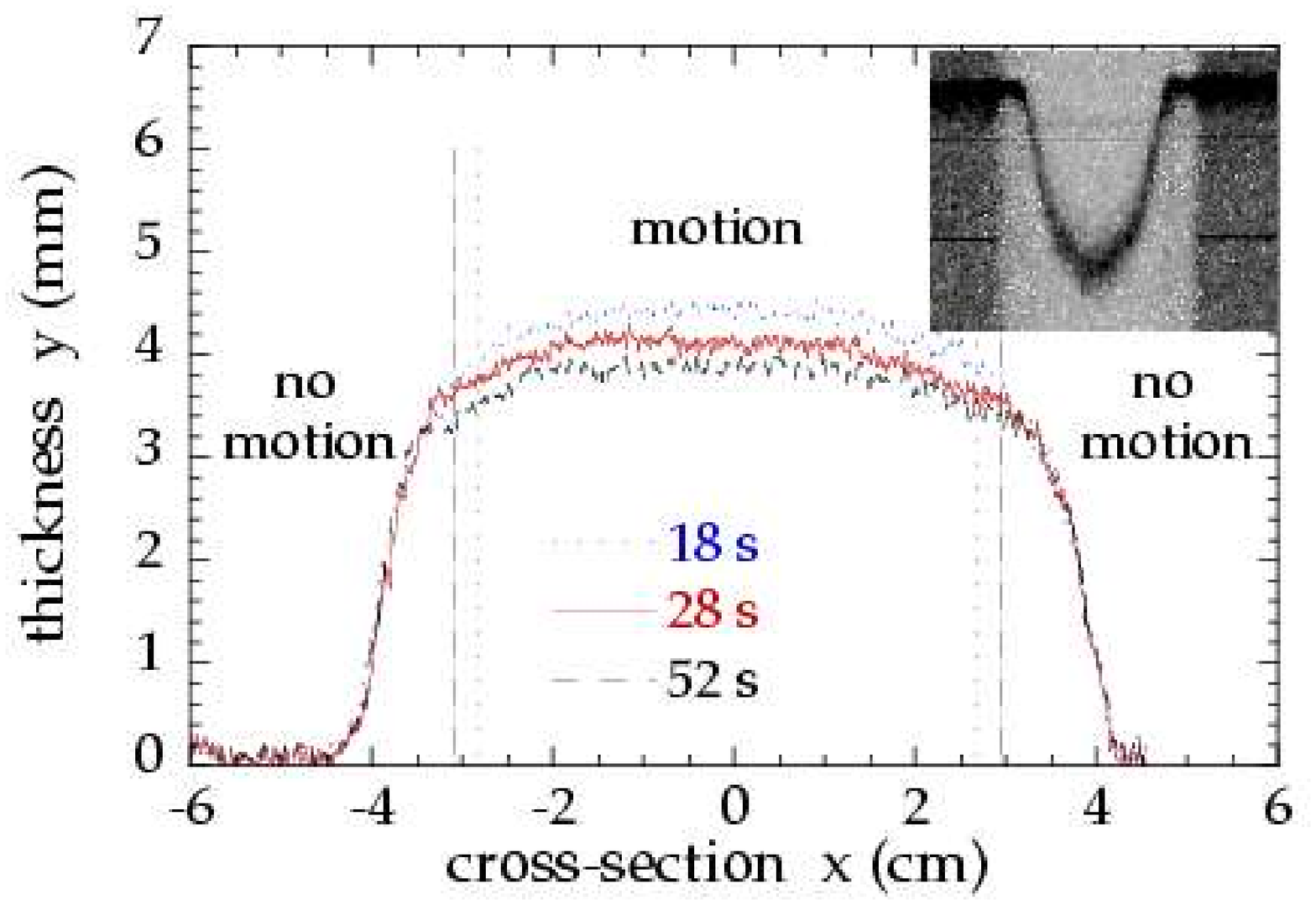}
\includegraphics[width=7.5cm]{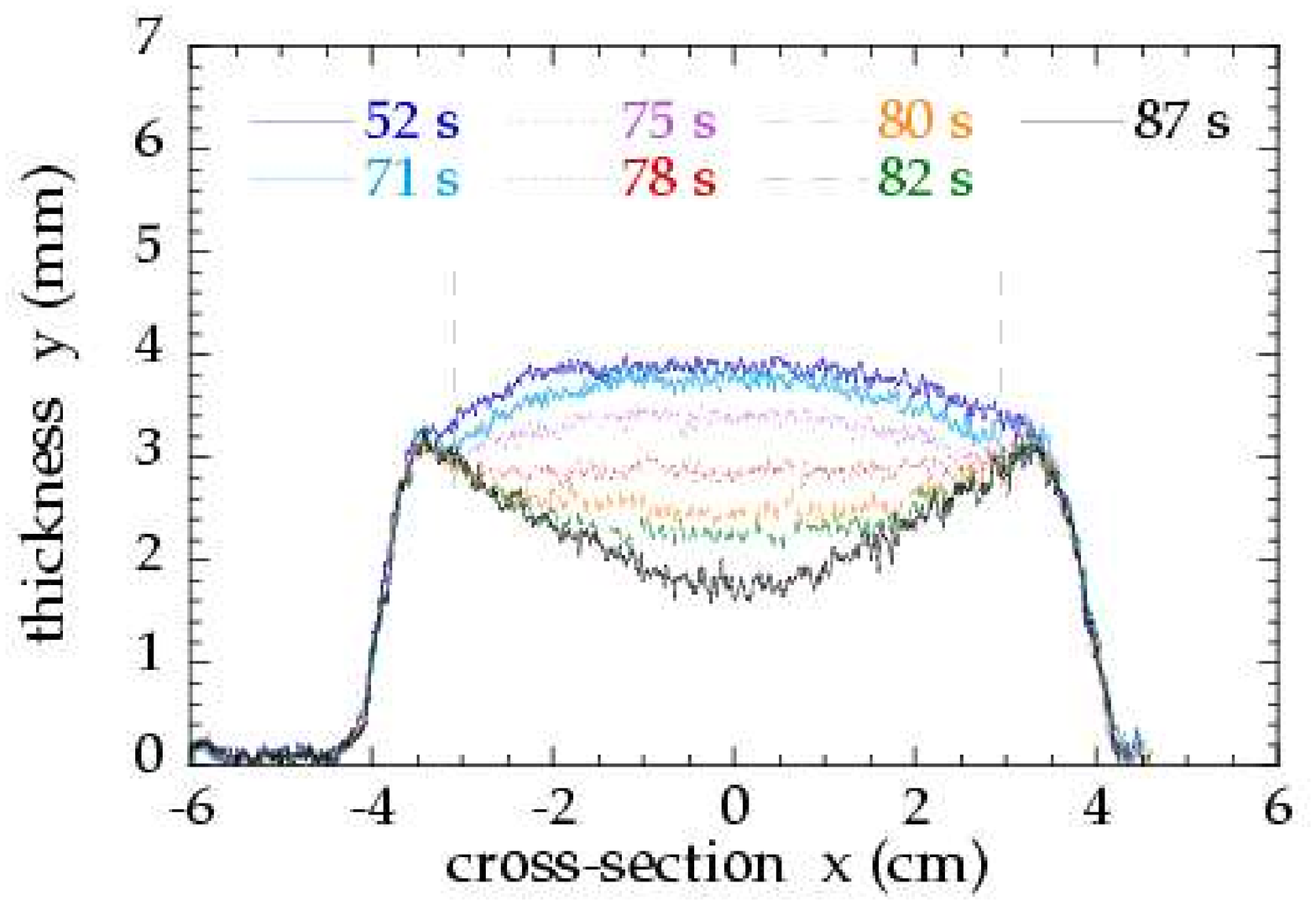}
\caption{(top) Arrival of the front of the flow at the chosen location: both height and width increase with time. X-axis is taken perpendicularly to the flow direction. (middle) Stable flow with constant width. The slight decrease of the height corresponds to a longitudinal convex shape of the front. The two lines limit the area where there is a surface motion: lateral static zones exist on each border of the flow (slightly wider at 18 s). Inset: a line of dark particles poured instantaneously on the surface of the flow shows the two static lateral borders and the surface velocity profile. (bottom) After stopping the supply, the central channel partly empties: the final morphology is high lateral lev\'ees with a low central channel. Here, the channel is not flat because of the `small' flux used. The lev\'ees record the height reached by the flow in the lateral static zones (beads: 300-400 $\mu$m, roughness: 425-600 $\mu$m, flux: 7 g/s, slope: 25$^\circ$)}
\label{frontvidange}
\end{figure}

The evolution with time of the flow morphology has been followed at a distance chosen such that the flow has reached its stable configuration. A `small' flux (see section \ref{reg}) is used for technical reasons. Three distinct phases occur (fig. \ref{frontvidange}):\newline
- First, the front of the flow arrives at the chosen distance: the height and the width of the cross-section increase until a stable profile is reached.\newline
- Second, the profile does not evolve. After a small decrease of the height corresponding to a slight longitudinal convex shape of the front, width and height are constant with time, at the chosen distance. They are also constant along the flow direction. We observe lateral static zones on each border of the flow: pictures of the laser lines taken with a long exposure time are blurred where beads are moving, and not blurred where there is no surface motion (limits drawn in fig. \ref{frontvidange}). The static zones can also be seen when a line of dark particles is poured instantaneously on the surface of the flow (inset). The line, initially perpendicular to the flow direction, evolves showing the surface velocity profile: the dark line is deviated neither on the plane, nor on the lateral borders of the flow.

\begin{figure}[htbp]
\includegraphics[width=7.5cm]{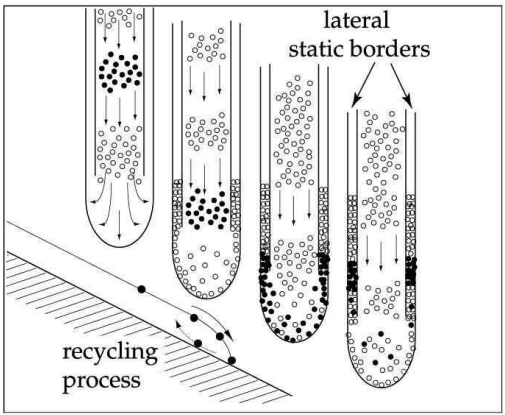}
\caption{Schematic representation of particle trajectories at the flow surface: particles arriving at the front, either go to build the lateral static borders at the rear of the front, or are recycled in the front when close to the center of the flow. Consequently, the lev\'ees record, down along the slope, the temporal evolution of the front composition.}
\label{traject}
\end{figure}

A small volume of colored particles was also suddenly released from the funnel and their trajectories were followed on the flow surface. Roughly, particles reaching the center of the front are re-injected into its rear, while particles reaching the lateral borders of the front become static, building the lev\'ees (fig. \ref{traject}). Thus, the composition of the static zones records,  down along the slope, the temporal evolution of the composition of the front.
A thin line of colored particles was also placed on the plane, on the future path of the flow. These particles reappeared on the surface of the flow, at the rear of the front, showing that the recycling process involves the whole thickness of the flow, and that the basal part of the flow is not static. 
\newline
- Finally, as the supply stops, the height in the channel decreases due to the downward flow (fig. \ref{frontvidange}), while the height of the static borders remains unchanged. Consequently, the central channel becomes lower than the borders, leading to the lev\'ee/channel morphology.  More accurately, the static borders partly collapse into the channel when it empties. The summits of the lev\'ees are slightly less high than the maximum height of static zones, and also a little further apart than the limits of static zones. The lev\'ees record the maximum height reached by the flow at the future location of their summits. They are thus linked to the height of lateral static zones just behind the front, and consequently linked to the front height. 
Here, we can associate the height of lev\'ees with the height of the flow, because the longitudinal convex shape of the front is small. But, when the front has a stronger convexity (case of some polydisperse media), it would be more accurate to associate the height of lev\'ees and the maximal thickness of the front.\newline

Consequently, the formation of lev\'ees is associated with the existence of lateral static zones and the partial emptying of the central portion of the flow. This is compatible with the precise observations and interpretation given by \cite{Rowley}.

\begin{figure}[htbp]
\includegraphics[width=7.5cm]{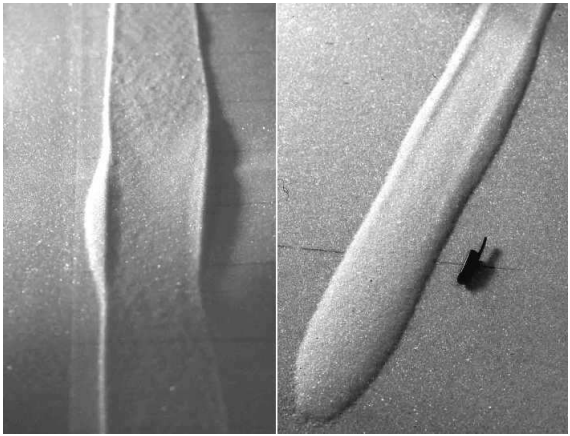}
\caption{(left) A local variation of the slope from 23.2$^\circ$ to 29$^\circ$, on a mean slope of 26$^\circ$, induces a local enhancement of the morphology; (right) after the supply stops, the front of deposit exhibits a rounded shape turning progressively into the lev\'ee/channel morphology at the back.}
\label{frontStop}
\end{figure}

When supply stops, the flow continues to propagate on a short distance, using the flux coming from the partial emptying of the back of the deposit. The deposit exhibits a smooth transition between a rounded front and the lev\'ee/channel morphology at the back, of which the deposit mainly consists (fig. \ref{frontStop}, right). This morphology is comparable to those observed in the field \cite{Rowley}. It has also been experimentally observed in the failure of a bump of sand \cite{McDonald}.

				\subsection{Evolution of the lev\'ee/channel morphology}

		\subsubsection{Evolution with the angle of the slope}
		
\begin{figure}[htbp]
\center
\includegraphics[width=7.5cm]{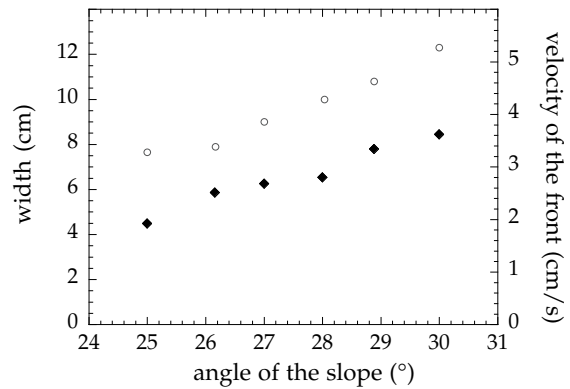}
\caption{({\large$\circ$}) width of flow and deposit, and ({$\blacklozenge$}) mean flow velocity (i.e. velocity of the front) both increase with the angle of the slope (flux: 7.1 g/s, beads: 300-400 $\mu$m, roughness: 425-600 $\mu$m).}
\label{VLpente}
\end{figure}

\begin{figure}[htbp]
\includegraphics[width=7.5cm]{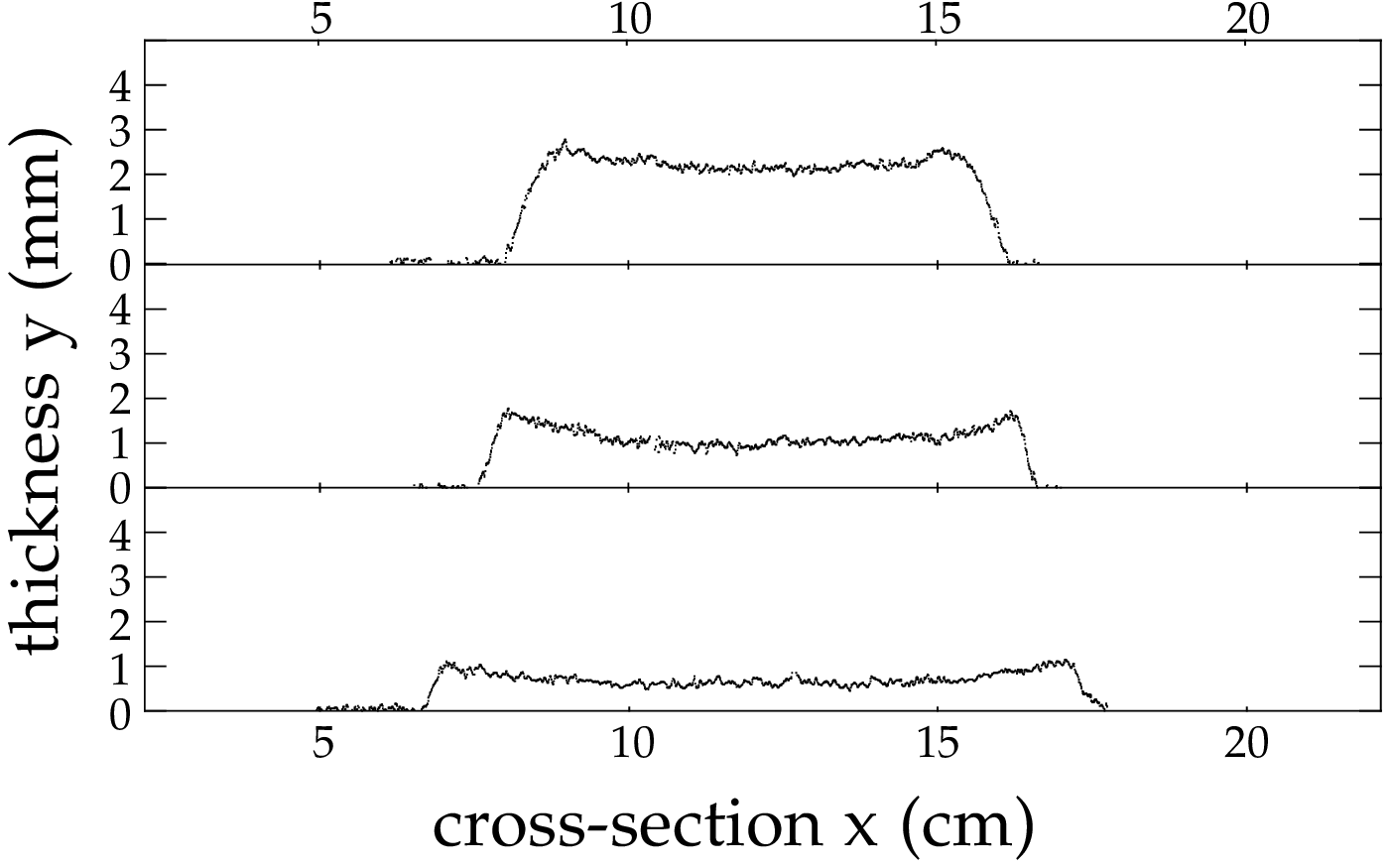}
\includegraphics[width=7.5cm]{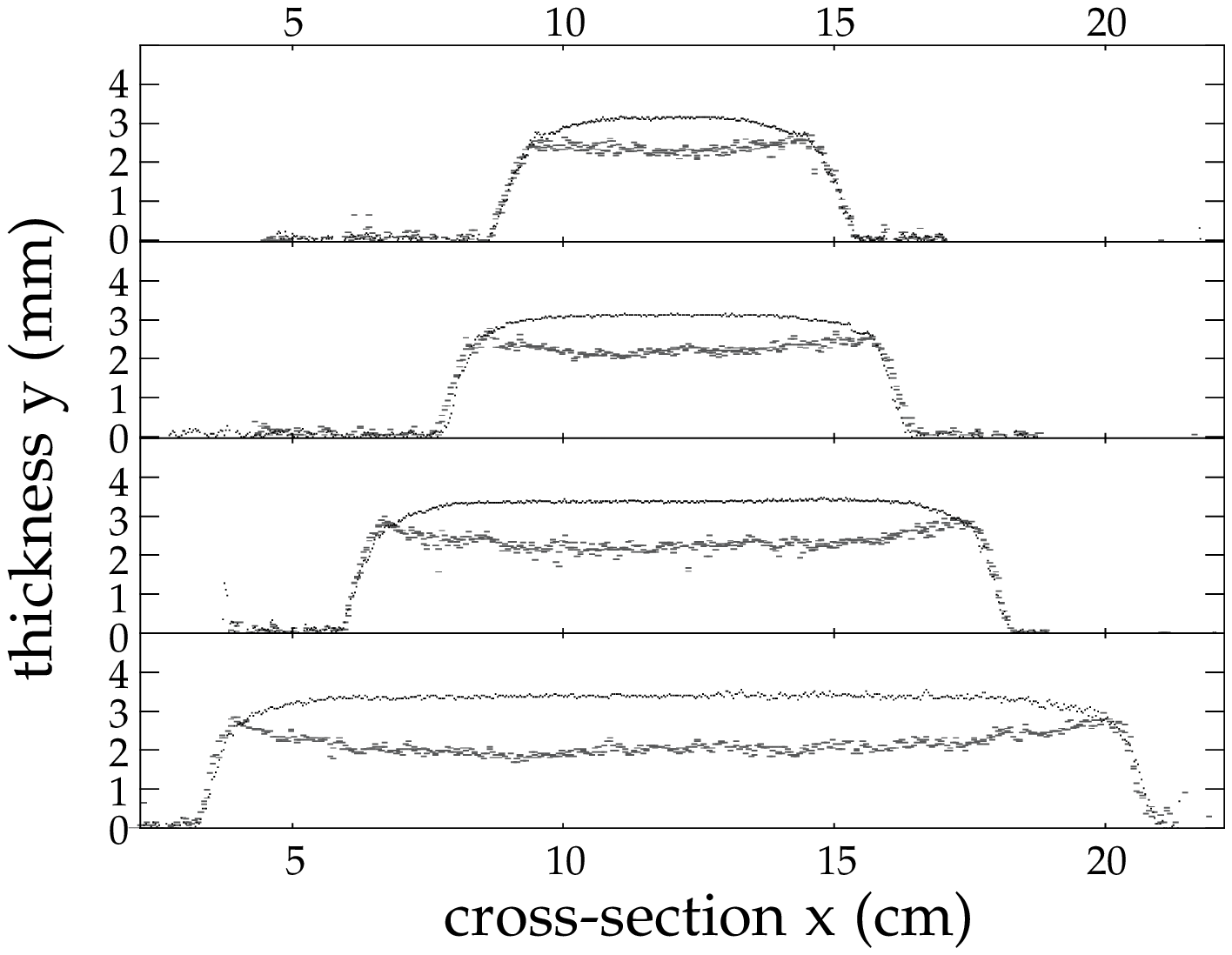}
\caption{(top) Increasing the angle of the slope enhances the morphology of the deposit. From top to bottom: 25$^\circ$, 27$^\circ$, and 29$^\circ$ (flux: 7.1 g/s). (bottom) Evolution of the morphology of a flow and its deposit for different fluxes, from top to bottom: 3.7, 8, 17.5, and 38 g/s (slope: 25$^\circ$). The width of the lev\'ees varies slightly, although the width of the deposit varies a lot (beads: 300-400 $\mu$m, roughness: 425-600 $\mu$m).}
\label{profilspf}
\end{figure}

The morphological aspect ratio varies when the inclination of the plane $\theta$ changes (fig. \ref{profilspf}, top). Because both lev\'ees height, $h_{lev\acute{e} e}$, and channel thickness, $h_{channel}$ decrease with $\theta$, $h_{channel}$ being inferior to $h_{lev\acute{e} e}$, the morphology is enhanced for  steep slopes, as observed in the field \cite{wilson}. Simultaneously, width and velocity increase with $\theta$ (fig. \ref{VLpente}).

\begin{figure}[htbp]
\includegraphics[width=7.5cm]{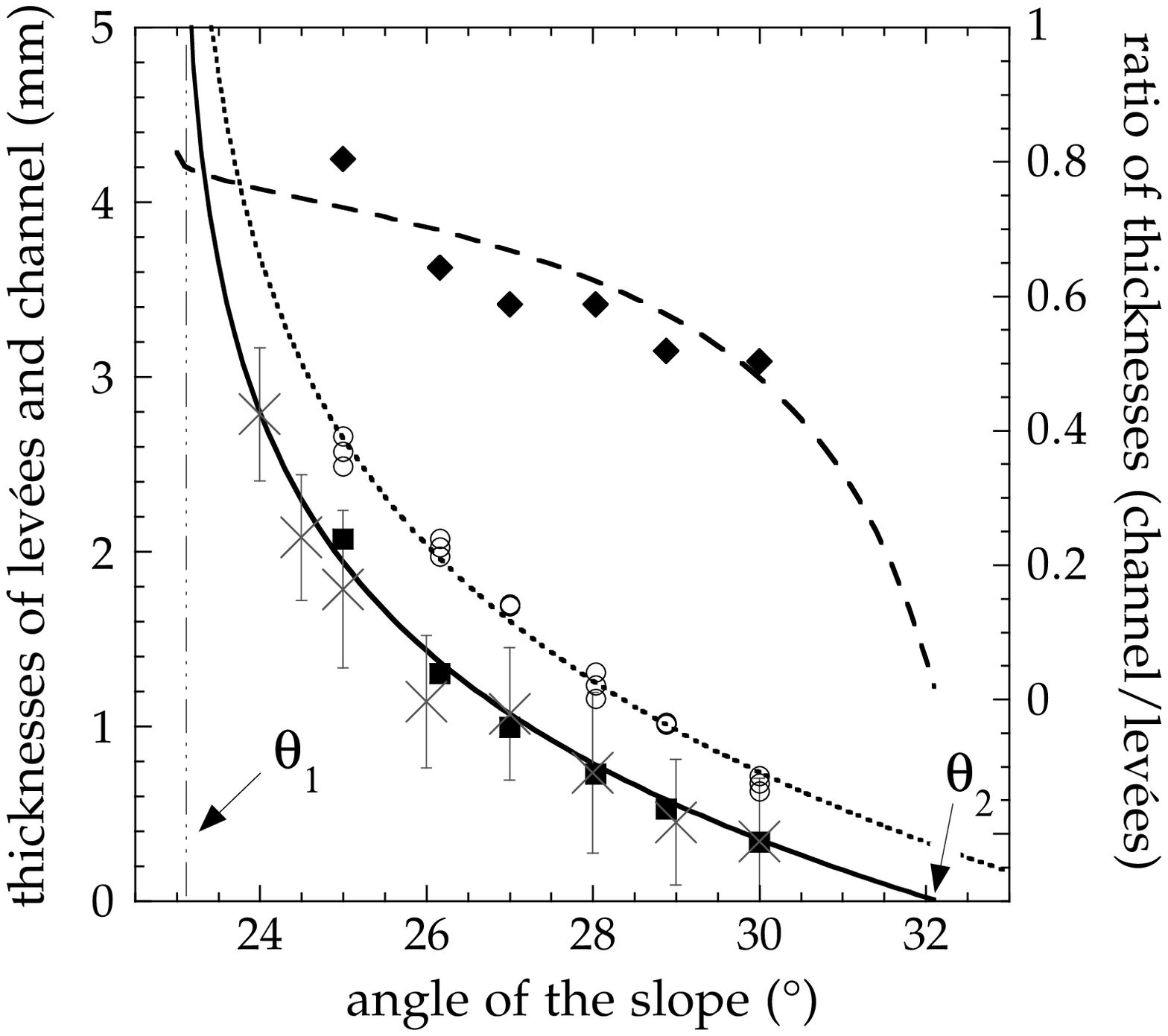}
\caption{Thickness of deposit: ({\Large$\circ$}) $h_{lev\acute{e} e}$, and ($\blacksquare$) $h_{channel}$. Fits come from equation (\ref{fit}) with the same values of $\theta_1$ and $\theta_2$ according to: (---) ``$h_{stop}$ curve", and (....) ``$h_{stop}+constant$" curve.  ($\blacklozenge$) the $h_{chennal}$/$h_{lev\acute{e} e}$ ratio and (- -) the fit obtained from the ratio of both fits show the decrease observed in the field. ({\large$\times$}) Thickness of the deposit of wide laterally confined flows, with error bars. Thicknesses of the deposit of both an unconfined flow, and a confined flow are equal (flux: 7 g/s for confined flows, and large enough to cover the whole plane for confined flows, beads: 300-400 $\mu$m, roughness: 425-600 $\mu$m).}
\label{morphopente}
\end{figure}

The dynamics of confined dense granular flows down rough inclines show that in a range of inclinations ($\theta_1$, $\theta_2$), a uniform layer of deposit remains after a flow has passed \cite{olivier, douady}. 
The deposit thickness, named $h_{stop}$, is independent of the velocity and of 
the thickness of the flow $h$, but depends on $\theta$ (fig. \ref{morphopente}). The thickness $h_{stop}$ is interpreted as a characteristic length on which stresses generated by the friction with the substratum are transmitted in the granular matter. It is linked to the precise characteristics of the plane roughness and of the particle size \cite{celine}.
$h_{stop}$ is one possible vertical length scale for the system \cite{olivier}.

Previous authors found that 
there is also a maximum thickness of material which can stay on an inclined plane without moving, $h_{start}$ \cite{oliviercalotte, douady}; $h_{start}$ depends on $\theta$, generating a similar curve, shifted towards large angles ($h_{start}>h_{stop}$).
A layer of granular matter with a thickness smaller than $h_{stop}$ is stable and never flows. If 
its thickness (or $\theta$) increases and reaches $h_{start}(\theta)$, the layer destabilises and flows; the produced deposit having the $h_{stop}$ thickness. For a thickness between $h_{stop}$ and $h_{start}$, the granular layer can be stable or unstable depending on the initial conditions (hysteresis). 

The solid line in the figure \ref{morphopente} is a fit for $h_{stop}(\theta)$ established by Pouliquen \cite{olivier} where $C$ is an adaptable parameter and $d$ the size of the beads:
\begin{equation}
\tan\theta=\tan\theta_{1}+\left(\tan\theta_{2}-\tan\theta_{1}\right)\exp\left(-\frac{h_{stop}}{Cd}\right).
\label {fit}
\end{equation}

In our experiments, $h_{channel}$ and $h_{lev\acute{e} e}$ show a decreasing trend compatible with a `$h_{stop}(\theta)$' type curve, although $h_{lev\acute{e} e}$ could be an `$h_{stop}(\theta)+constant$' type curve (fig. \ref{morphopente}). However, we do not have enough data to check that point. This shows that the thickness of deposit in the channel is linked to the stopping process,  and is independent of height and velocity of the flow.

To confirm that $h_{channel}$ corresponds to the $h_{stop}$ parameter defined by Pouliquen \cite{olivier}, we measured the thickness of our material lying on our plane after a wide confined flow has passed. The flows were made using the 40 cm gate, sufficiently opened to obtain a thick sheet flow, which rapidly reaches the edges of the plane, giving a confined flow and a flat deposit. The measurements
are less precise because the scale cannot be on the same picture as the deposit. Nevertheless, data on both thicknesses are in good agreement (fig. \ref{morphopente}).

In the field, the lev\'ee/channel morphology has been described by the ratio $h_{channel}$/$h_{lev\acute{e} e}$. This ratio is sensitive to errors: we fit both sets of data, and calculate the ratio of each fit (fig. \ref{morphopente}). 
The decrease of the ratio with $\theta$ is compatible with observations on pyroclastic deposits: a ratio of 0.9 for `low' slopes, and of 0.5 for `steep' slopes on 1980 Mount St. Helens deposits \cite{wilson}. 
Moreover, the morphology adapts very rapidly to a local variation of the slope, leading to a local deformation of the deposit (fig. \ref{frontStop}, left), which has been observed in the field \cite{wilson}.

		\subsubsection{Evolution of the morphology with the flux}

A flux increase induces an increase of the flow velocity and width (fig. \ref{largeurdist}). At the same time, the thickness adapts. Figure \ref{profilspf} (bottom) shows a comparison of flow and deposit cross-sections for different fluxes. First, except for very narrow flows, the morphology of flow and deposit is mainly flat. This is appropriate to define $h$ as the thickness of flow and $h_{channel}$ as the thickness of deposit in the channel. For narrow flows (fig. \ref{frontvidange}), $h$ is taken as the maximum thickness of the flow, and $h_{channel}$ as the minimum deposit thickness. Second, the width of the lev\'ees is almost the same, independently of the width of the flow, i.e. independently of the flux. Consequently, the aspect ratio changes as the flux varies. This morphological change is different from those induced by a slope variation (fig. \ref{profilspf}, top), giving a way of differentiating the effects of flux and of topography in the field.

		\subsection{Flow regimes and associated deposit morphologies}\label{reg}

The second step of this study consists in linking  the flow dynamics and the deposit morphology. A qualitative link is presented first, then a quantitative relation is determined for one regime and, finally, the influence of polydispersity of materials is discussed.  

		\subsubsection{Description of the regimes}

\begin{figure}[htbp]
\begin{minipage}[b]{0.0\linewidth}
\center
\includegraphics[width=3.269cm]{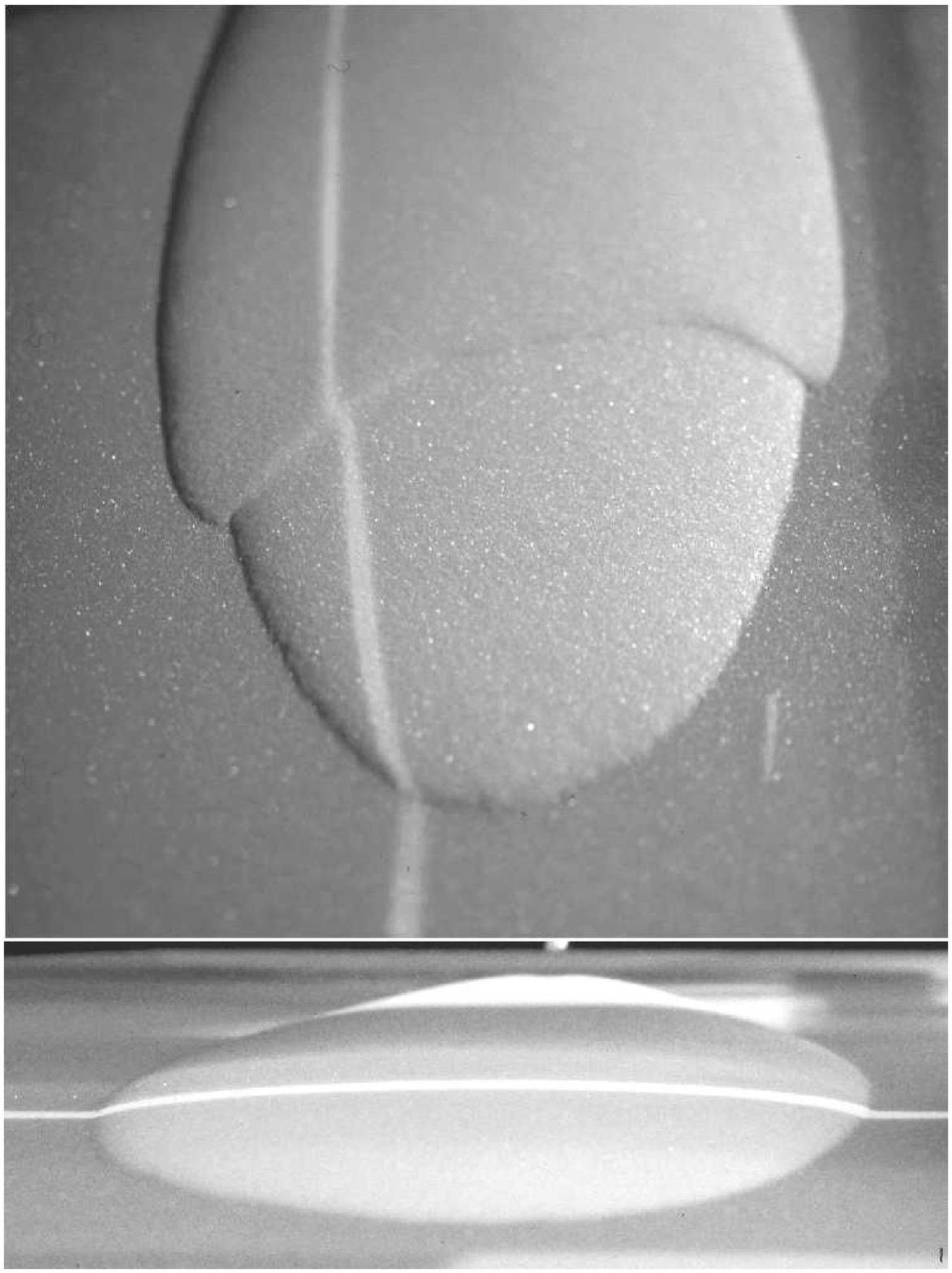}
\end{minipage}
\hfill
\begin{minipage}[b]{0.600\linewidth}
\center
\includegraphics[width=4.231cm]{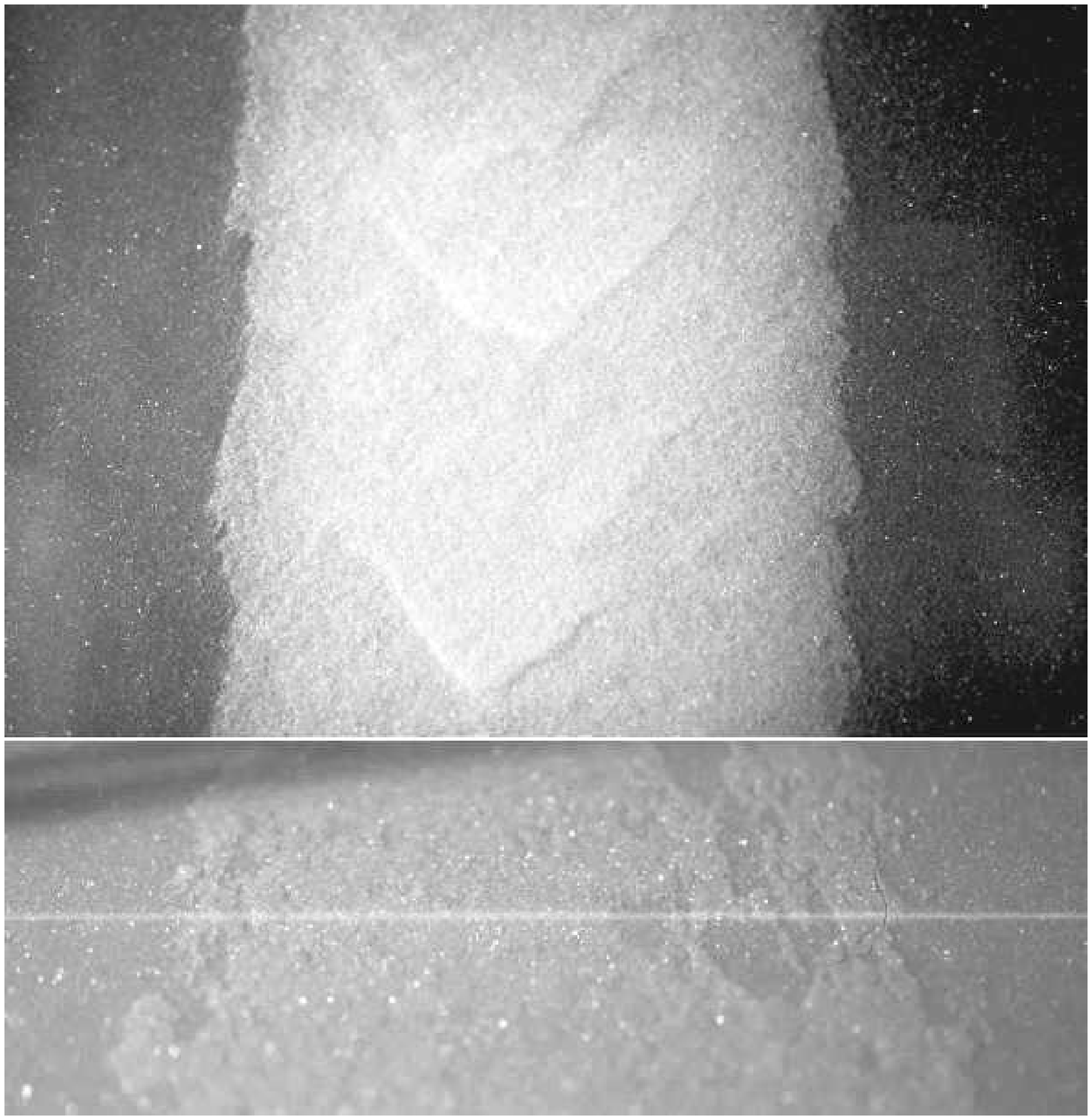}
\end{minipage}
\caption{Two other types of (top) flow and (bottom) associated deposit:
(left) intermittent flow regime and rounded shape deposit (for low flux and/or slope), and (right) roll-wave regime and thin flat deposit (for high flux and slope). The line is the projection of the laser sheet either perpendicular or parallel to the flow direction.}
\label{photoregimes}
\end{figure}

\begin{figure}[htbp]
\includegraphics[width=7.5cm]{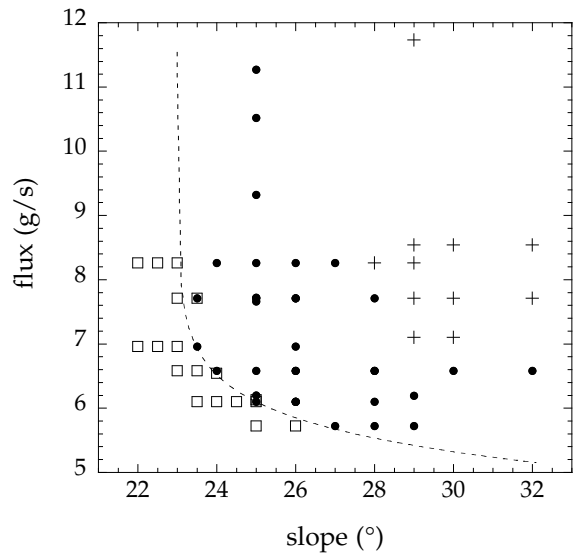}
\caption{Flow regime fields: ({\footnotesize{$\square$}}) intermittent flow regime (see fig. \ref{photoregimes}, left), ({\bf +}) roll-wave flow regime (see fig. \ref{photoregimes}, right), and ({\Large{$\bullet$}}) steady state flow regime (see fig. \ref{photolevee}). The curve comes from the $h_{stop}(\theta)$ curve ((---) in fig. \ref{morphopente}), adapted with a linear relation between $h_{stop}$ and the flux.}
\label{regimes}
\end{figure}

This study is focused on dense frictional flows. The collisional regime has not been studied, although it is observed for higher slopes.
Several flow regimes happen depending on flux and inclination of the plane. The volume fraction of beads is between a random loose and a random close packing of beads (0.56-0.64\%) \cite{onoda}, and if there is any difference between the regimes, they would be to small to be measured. Each type of flow has its own dynamics, and produces a particular type of deposit (figs. \ref{photolevee} and \ref{photoregimes}). Three regimes can be defined, depending on values of slope and flux (fig. \ref{regimes}). In this section, experiments involve particles with narrow size distributions (215-224, 150-250 or 300-400 $\mu$m), on 425-600 $\mu$m roughness. This has been done to obtain various values of the flux. Results are consistent even if beads of different sizes are used on the same substratum, because the size changes only slightly. A few test experiments have been done with rough painted beads, and with other bead sizes. They show a dependency of the limits of the regime fields on the characteristics of the particles. \newline

-First, for low flux and/or low slope, an intermittent flow regime is observed (figs. \ref{photoregimes}, left and \ref{regimes}). A rounded shaped flow spreads down the slope, then  stops. As it is continuously supplied with granular matter at the back, a second flow starts  a few seconds later from the  accumulation zone, and covers the previous one, propagating the whole deposit a little further, and then it stops again. In the figure \ref{photoregimes} (top, left), the front is static, while the back is moving ahead (blurred area). The flow is consequently intermittent, with a front velocity alternating between 0 and a positive value, which might be similar to features observed during Montserrat eruptions \cite{cole98}. No measurement of frequency has been done, which seems to depend on both flux and slope. When the deposit is long enough, several successive avalanches can propagate simultaneously. The deposit has a rounded shape (fig. \ref{photoregimes}, bottom, left). Sometimes, when the feeding has been stopped early, it exhibits a step, the provided volume being not large enough for a total covering.\newline
-Second, for a given intermediate range of flux and inclination, a steady state flow is observed (figs. \ref{photolevee}, \ref{regimes} and section \ref{levee}). The flow is finger-shaped, with constant width, height, and velocity. The deposit displays the lev\'ee/channel morphology. For this regime, a `small' flux corresponds to conditions close to the limit between intermittent and steady state regimes.\newline
-Finally, for high flux and slope, a roll-wave regime occurs (figs. \ref{photoregimes}, right and \ref{regimes}). The flows are thin and finger-shaped with parallel lateral borders which are not very-well defined. Small waves develop at the surface of  the flow and rapidly reach its front. The deposit is either flat and as thin as 1-2 layers of beads, or non-existent.

These three regimes are closely related to those observed for wide confined flows \cite{olivier, yoel}, when making an equivalence between the flux and the flow thickness: for low angle and/or thickness, no flow occurs; for intermediate values, flows are in a steady state regime; for high angle and thickness, instabilities appear, which are predicted by the theory \cite{shyam, yoel}. Using sand, these waves appear as soon as there is a flow \cite{yoel}, and could be associated with the lev\'ee/channel morphology \cite{McDonald}.  

These regimes obey different dynamics. It is important to know in which regime the flow responsible for a deposit has propagated, before any attempt to calculate its velocity. Fortunately, the deposits have different morphologies, and the flow regime can be deduced from qualitative observations of the deposit. 
Note that the determination of the regime is clear when the lev\'ee/channel morphology is observed. But when the deposit is rounded, it can be either the result of an intermittent regime, or the result of the propagation of the flow within the steady state regime, the observation being made too close to the front to see the emptying of the back of the deposit (fig. \ref{frontStop}, right).

		\subsubsection{Interpretation using the thickness $h_{stop}(\theta)$}

It is possible to interpret the regimes using the thickness $h_{stop}(\theta)$ remaining after a wide confined flow has passed (fig. \ref{morphopente} and eq. \ref{fit}). Two angles are involved: for $\theta_1$, $h_{stop}$ tends towards infinity, while above $\theta_2$, it is equal to zero. This means that below $\theta_1$, no uniform layer of granular matter can be thick enough to flow, and that above $\theta_2$, no deposit remains (even if some matter may rest on a slope slightly above $\theta_2$, when placed in a static state on the plane). The dashed curve (fig. \ref{regimes}) comes from the $h_{stop}(\theta)$ of the figure \ref{morphopente}, assuming that the flux and $h_{stop}$ are linked by a linear relation. That shows that the limit between steady state and intermittent regimes, and the $h_{stop}(\theta)$ curve both go through a compatible range of angles ($\theta_1$, $\theta_2$).\newline
-Under the $h_{stop}(\theta)$ curve, a wide confined granular layer does not flow \cite{olivier}. When granular matter is released from a funnel, it cannot initiate a flowing layer with a local constant thickness, either because the slope is below $\theta_1$, or because the flux is too low to feed a flow thicker than $h_{stop}$. It constitutes a heap till its free surface reaches an angle above the stability angle. Then, the heap collapses and the granular matter flows in a non-symmetrical way on the slope. The intermittent regime is the result of successive avalanches building a `pile' on a steep slope, because a uniform layer would not flow.\newline
-Above the $h_{stop}(\theta)$ curve, the amount of released matter is sufficient to feed a layer which is locally thicker than $h_{stop}$: thus the steady state regime occurs.\newline 
-For angles {\it slightly} above $\theta_2$, the deposit consists of two lev\'ees aside an empty channel, because for every angle, the thickness of the lev\'ees is larger than those of the channel (fig. \ref{morphopente}).\newline
-Above $\theta_2$, the particles flow down and no deposit stays on the plane. Extrapolating our results, it seems that both the steady state and roll-wave regime can occur. It is also very likely that the collisional regime occurs.

\begin{figure}[htbp]
\center
\includegraphics[width=7.5cm]{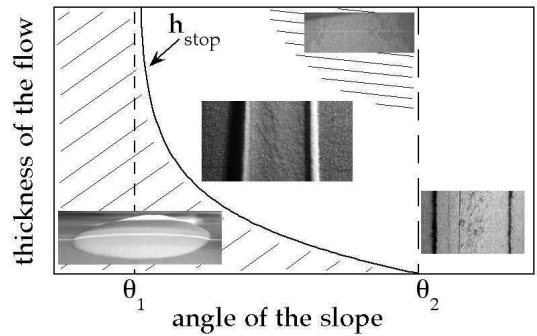}
\caption{Location of the deposit types, shown by the inset photographs, compared to the $h_{stop}(\theta)$ curve. For flow conditions below the curve, the deposit has a rounded shape. The lev\'ee/channel morphology exists above the curve except when instabilities occur (high angle and thickness). `Lev\'ees with an empty channel' can be observed for  angles slightly above $\theta_2$ (dark beads for this inset). For higher angles, there is no deposit either in the lev\'ees, or in the channel. Consequently, the lev\'ee/channel morphology, with some matter remaining in the channel, occurs only between the $\theta_1$ and $\theta_2$ angles.}
\label{regimesHstop}
\end{figure}

Consequently, the potential occurrence of the different types of deposits can be located in comparison with the $h_{stop}(\theta)$ curve (fig. \ref{regimesHstop}). 
\newline-For slopes smaller than $\theta_1$, the deposit is thick and rounded in shape.\newline-For slopes between $\theta_1$ and $\theta_2$, the three regimes can occur. The deposit displays the lev\'ee/channel morphology except close to the front, or except if the flux is too low or too high. Very close to $\theta_2$, only a few particles stay on the plane ($h_{stop}<d$), corresponding to lag deposits observed on steep slopes \cite{Rowley}. \newline-For slopes greater than $\theta_2$, there is no deposit (at least in the channel), as it is observed in the field \cite{davies, Yamamoto, Calder}. For angles slightly above $\theta_2$, the ``lev\'ees with empty channel" morphology exists (fig. \ref{regimesHstop}, right inset).

We conclude from these results that if the lev\'ee/channel morphology is observed on a deposit, there is no doubt that the associated flow was in a steady state frictional regime, and that the angle of the slope was between $\theta_1$ and $\theta_2$. We made observations on the south-east lobes of the Lascar (Chile) 1993 deposits, which gives approximate values for $\theta_1$ and $\theta_2$ of 6$^\circ$ and 14$^\circ$. These values correspond to erosion/deposition behaviors observed for slopes respectively equal to 14-16$^\circ$ and 4-6$^\circ$ at Montserrat \cite{cole98}. Both angles are small, probably due to the large values of the ratio between substratum roughness and particle sizes \cite{celine}.  

Note that, for a given erupted volume, knowing $h_{stop}(\theta)$ is not enough to calculate the longitudinal deposit extension (except, if the flow in confined in a valley): it could be long and narrow (small flux) or short and wide (large flux).

				\subsection{From deposit morphology to quantitative information on flow dynamics}

There is a good analogy between confined and unconfined flows, confirmed by the existence of analogous regimes \cite{yoel} and we are confident to try successful methods used for dynamics of confined flows for the interpretation of dynamics of unconfined flows. 
The challenge is to obtain some quantitative values for the dynamics (flux, velocity, height of the flow) only from measurements performed on the deposit.  

\begin{figure}[htbp]
\includegraphics[width=7.5cm]{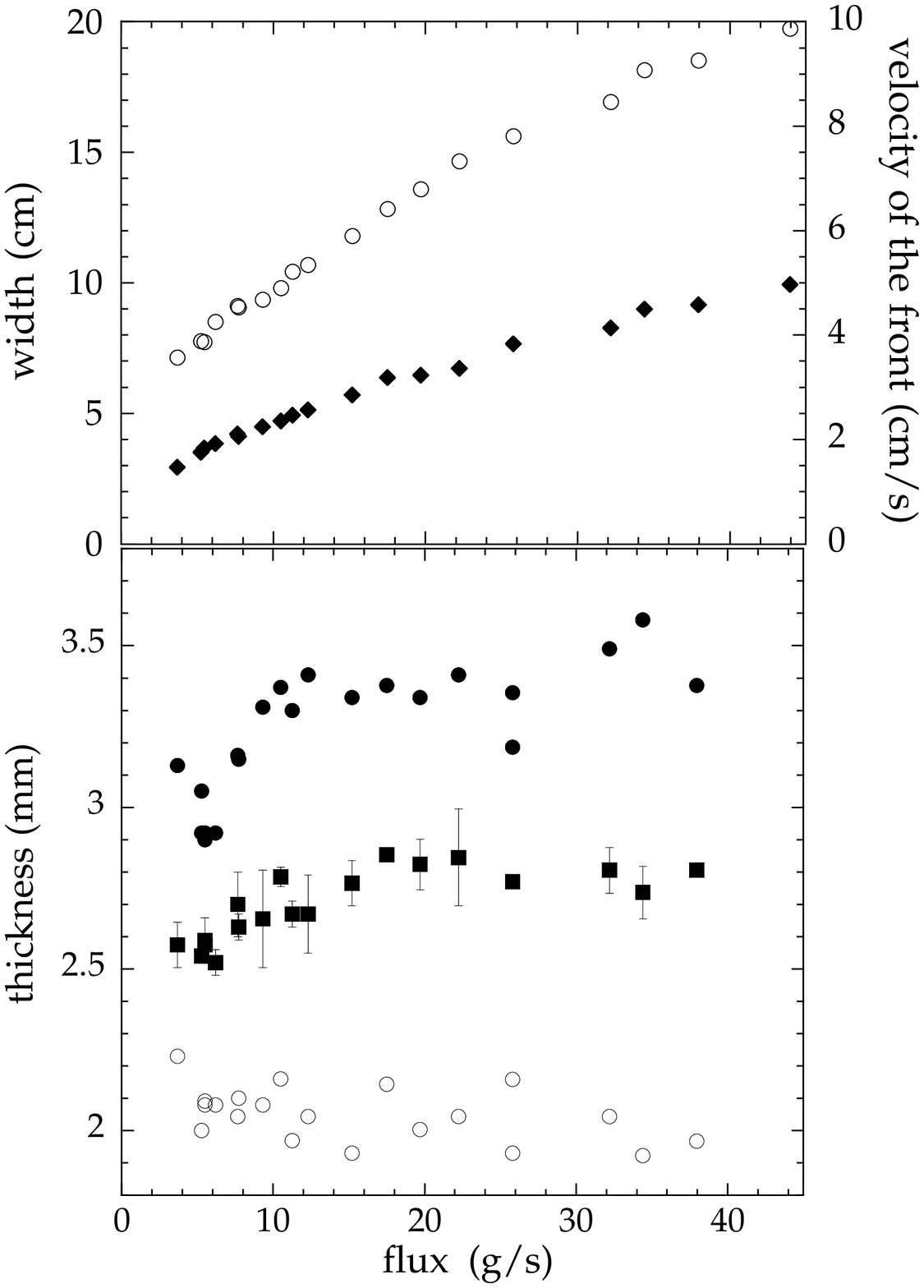}
\caption{(top) ({\large$\circ$}) Width of flow and deposit, and ({$\blacklozenge$}) mean flow velocity (i.e. velocity of the front) both increase with the flux. (bottom) Thickness of ({\large$\bullet$}) flow and ({\tiny$\blacksquare$}) lev\'ees increases slightly with the flux in a similar way. Error bars come from the difference between both lev\'ees heights in each deposit. ({\large$\circ$}) The thickness of material  in the channel of the deposit almost does not depend on the flux (beads: 300-400 $\mu$m, roughness: 425-600 $\mu$m, slope: 25$^\circ$).}
\label{toutfctflux}
\end{figure}

We focus on the lev\'ee/channel morphology because the particular shape of the deposit can be related without ambiguity to the flow regime.
In this regime, the influence of the flux has been tested. It appears that the larger the flux, the faster the front of the flow and the wider the flow (and the deposit) (fig. \ref{toutfctflux}). As expected \cite{olivier, douady}, $h_{channel}$ contains no information on dynamics, which is not the case for $h_{lev\acute{e} e}$ and $h$ (fig. \ref{toutfctflux}). Note that the variation of $h_{lev\acute{e} e}$ with the flux is incompatible with a 2-dimensional calculation of the dynamics of an infinite layer of fluid having a Bingham rheology, which would predict that the thickness of the static zone is determined by the yield stress and should depend only on the slope. 
The lev\'ee height is linked to $h$, through the formation of lateral static zones, and to $h_{channel}$, through the partial collapse.  Because the $h_{channel}$ thickness is constant (or very slightly decreasing) with the flux, the increase of $h$ and $h_{lev\acute{e} e}$ with the flux are similar. More accurately, the slightly larger value of $h_{channel}$ for small flux may be due to an arching process between the static borders during the collapse \cite{ammi, Mills, Luding}. Because the associated flow is narrow, the short distance between the static zones prevents the complete emptying until the $h_{stop}$ thickness of a wide flow deposit is reached. It would have been more accurate to keep the notation `$h_{stop}$' for deposit produced by a wide flow, and to say that the thickness in the channel, $h_{channel}$ tends to $h_{stop}$ when $L$ tends towards infinity. But, except for very narrow flows, $h_{channel}=h_{stop}$, and we chose not to distinguish between these 2 variables. The morphology of the deposit then contains parameters linked to the flux ($h_{lev\acute{e} e}$, $L$) and parameters independent of the dynamics ($h_{stop}$).

For wide laterally confined flows down rough inclines, Pouliquen \cite{olivier} first established a relation between the mean flow velocity $v$ and the flow thickness $h$ (eq. \ref{vit}). As previously, $h_{stop}$ is the thickness of the uniform deposited layer. It is the vertical length scale of the system and used to normalize the flow thickness:
\begin{equation}
\frac{v}{\sqrt{g h}}=\beta \frac{h}{h_{stop}},
\label{vit}
\end{equation}
$g$ is the gravity, and $\beta$ a constant equal to 0.136 for glass beads flowing down a plane made rough by gluing other glass beads. Due to the normalization by $h_{stop}$, all flows composed of different sizes of glass beads, on different planes of this type, follow the same law. This type of correlation is also verified in our experiments on unconfined flows, giving a similar value for $\beta$. 

\begin{figure}[htbp]
\includegraphics[width=7.5cm]{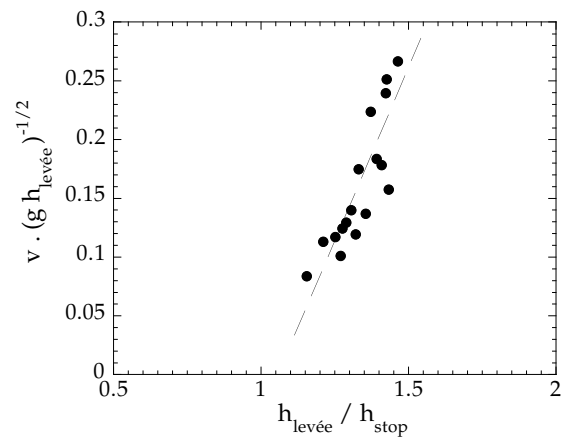}
\caption{Normalized correlation between velocity and thicknesses of the lev\'ees and the channel, for laterally unconfined granular flows and associated deposits (beads: 300-400 $\mu$m, roughness: 425-600 $\mu$m, slope: 25$^\circ$).}
\label{correl}
\end{figure}

Figure \ref{correl} shows the correlation obtained in the case of our experimental flows by using $h_{lev\acute{e} e}$ instead of $h$, and $h_{channel}$ for $h_{stop}$. As expected, the value of the constant $\beta$ is different from those previously obtained, being equal to $\beta$=0.5 (fig. \ref{correl}). The correlation exists, although the slope is high, showing that the correlation is of the same order of magnitude than the data variability. Unfortunately, due to the finite length of the experimental set-up, it was not possible to study the steady state regime for higher fluxes, and consequently to accurately determine this correlation.
Nevertheless, we deduce a relation allowing an estimate of the flow velocity only using deposit morphology: 
\begin{equation}
\frac{v}{\sqrt{g\, h_{lev\acute{e} e}}}=\beta\,\,\frac{h_{lev\acute{e} e}-h_{stop}}{h_{stop}}.
\label{vitdep}
\end{equation}

Its extension to natural systems is not direct since the value of the constant $\beta$ for volcanic granular material is unknown and probably different (experiments using sand give $\beta$=0.65 \cite{yoel}). We are also aware of the deflation of the deposits, which is low in most cases (0.3-1\% for St Helens deposits \cite{Rowley} or 2-6\% for Unzen deposits \cite{unzen}), but can also be important  (30-50\% for St Helens deposits \cite{Rowley}) and from which the deposit measurements should be corrected before using this relation. Each deposit have to be corrected in a different way and it is not possible to obtain a general equation including the deflation effect. The deflation is not calculated in this paper, since, to our knowledge, no precise morphological natural data are available. Moreover, natural materials have a wide size distribution inducing segregation processes that are not taken into account in relation \ref{vitdep}, and whose influence will be shown in the next section.

				\subsection{Flows with a wide particle size distribution}\label{polyd}

\begin{figure}[htbp]
\includegraphics[width=7.5cm]{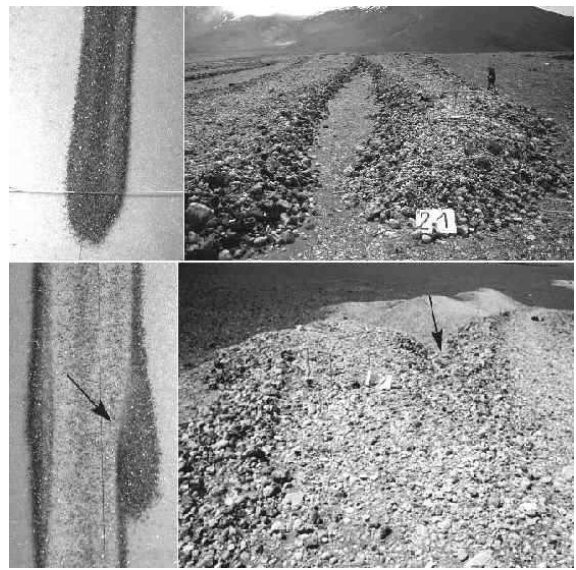} 
\caption{(top, left) Accumulation of large black particles in the front of laboratory granular flows, and (top, right) accumulation of large pumices in the front of the right lobe, in the Lascar pyroclastic flows deposits. Lobes fingering (bottom, left) in a glass beads flow deposit and (bottom, right) in a Lascar pyroclastic flow deposit (main flow going to the far right). The black arrow shows the point where the flow separates in two lobes. In the left picture, the small flow is particularly rounded in
 shape and still covered by large particles, because it did not propagate far enough to empty the back of the deposit. There is an accumulation of large black particles and of large pumices in the lev\'ees (see also fig. \ref{Lascar}, top).}
\label{frontdigit}
\end{figure}

During experiments involving wider size distributions, segregation features have been observed on flow and deposit. The large particles rapidly reach the surface of the flow and consequently accumulate at the front and in the static borders. In the deposit, the lev\'ees and the front are enriched in large particles while the  channel contains mainly small particles, as observed in the field (figs. \ref{Lascar}, \ref{frontdigit} and \cite{Calder}). 

Segregation is also responsible for fingering of the flows observed in the laboratory as well as in the field (black arrow in fig. \ref{frontdigit}, bottom). The jamming of a large particle when it reaches the substratum at the front of the flow induces the flow to separate into two distinct lobes \cite{olivierdigit}.

The influence of the size distribution on the velocity of the front and on the morphology of the deposit was also tested. It results from two processes: (1) contacts between particles in the flow are modified, i.e. the rheology changes; (2) the basal friction is strongly modified depending on which particle size is in contact with the plane. The friction depends on the size ratio between particles and substratum roughness. For monodisperse roughness and grain sizes, it has been shown that the friction is smaller for high and small ratios, than for ratios close to 1 \cite{celine}. For that reason, $h$ in equation (\ref{vit}) needs  a normalization by $h_{stop}$. Because of segregation, fine particles accumulate in the basal part of the flow, and eventually, fill the holes, smoothing the roughness. The plane thus becomes smoother for a large particle, but could be more rough for a small one. The friction between flow and substratum is mainly the result of the interaction between these small beads and the plane and could be reduced or enhanced depending on which system it is compared to (pure flow of small particles or pure flow of large ones), and depending on the ratios of particles and roughness sizes.

 \subsubsection{Bidisperse media}

The influence of fine particles has been observed in bidisperse flows involving a small fraction (respectively 5 and 10 weight\%) of 45-90 $\mu$m beads in 300-400 $\mu$m beads. For such sizes, the fines fill the holes of the plane constituted of 425-600 $\mu$m glued beads. The velocity increases compared to the flow of pure large particles (by a factor of, respectively, 1.4 and 1.9), and the morphology is strongly enhanced: the channel is almost empty between two high lev\'ees (see fig. \ref{regimesHstop}, right inset). 

On the same plane, each particle size is associated with a different quasi-parallel $h_{stop}(\theta)$ curve \cite{celine}. For these sizes and roughness, a model \cite{celine} predicts that the $h_{stop}(\theta)$ curve of the fines is shifted to small angles compared to the $h_{stop}(\theta)$ curve of the large ones. In bidisperse mixtures, segregation puts the fine particles at the bottom of the flow, and the thickness of the deposit in the channel is mainly imposed by the $h_{stop}(\theta)$ curve of the fine particles. But, pure large particle flow and mixture particle flow would roughly lead to similar lev\'ees height, lev\'ees being built at the rear of the front, mainly constituted of large particles in both cases. 
At the present angle (27$^\circ$), small particles fill in the channel during the flow, but the channel deposit is almost empty, because the slope is close to $\theta_2$ of the fines particles. On the contrary, the channel contains some matter for a flow of pure large particles, the present angle being far under $\theta_2$ of the large particles. Moreover, a small value of $h_{stop}$ corresponds to a low basal friction: the velocity of the bidisperse flow is higher than the velocity of a pure large particle flow. The shift of the $h_{stop}(\theta)$ curve for the small particles is responsible for enlarging the range of angles where the ``lev\'ees with empty channel" morphology occurs. This morphology has been observed at least from 25$^\circ$ to 30$^\circ$, and the associated lev\'ees are not necessarily small, which is not the case of monodisperse flows. 

   \subsubsection{Continuous polydisperse media}
Continuous polydisperse granular media have also been used. Successive mixtures (table I) have been made, corresponding to a progressive increase in the width of the size distribution.

\begin{table}[h]
\caption{Composition of the polydisperse mixtures: abundances are in weight\% (left column: size ranges in $\mu$m; top line: batch number used in the figures).}
\begin{center}
\begin{tabular}{|l|c|c|c|c|c|c|c|c|}
\hline
size/batch:&1&2&3&4&5&6&7&8\\
\hline
215-224&100& 0 & 0 &0 &0&0 &0&0\\
\hline
45-90&0&0&0&0& 0&1&4&7\\
\hline
70-110 &0& 0& 2&5&8&8&8&7\\
\hline
150-250&0& 100& 96&90&84&82&76&72\\
\hline
300-400&0& 0& 2&5&8&8&8&7\\
\hline
425-600&0&0&0&0& 0&1&4&7\\
\hline
\end{tabular}
\end{center}
\end{table}

\begin{figure}[htbp]
\includegraphics[width=7.5cm]{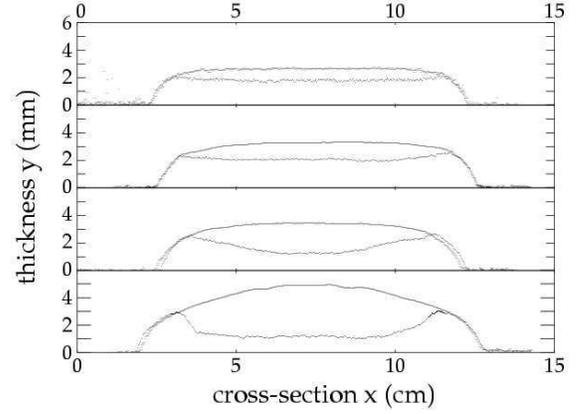}
\caption{Flow and deposit morphologies for an increasing polydispersity degree, from top to bottom: batches 2, 3, 4, and 8 (see table I for size distributions) (flux: 8.15 g/s, roughness: 425-600 $\mu$m, slope: 25$^\circ$).}
\label{poly}
\end{figure}

Morphologies of flows and deposits are enhanced when the size distribution is wider (fig. \ref{poly}). This trend quickly appears and is linked to the segregation process combined with the shift  of $h_{stop}$ curves when changing the smallest and the largest sizes in the distribution. Observations on deposits show that when the size distribution is increased, the particles of the channel become smaller and the particles of the lev\'ees larger.

\begin{figure}[htbp]
\includegraphics[width=7.5cm]{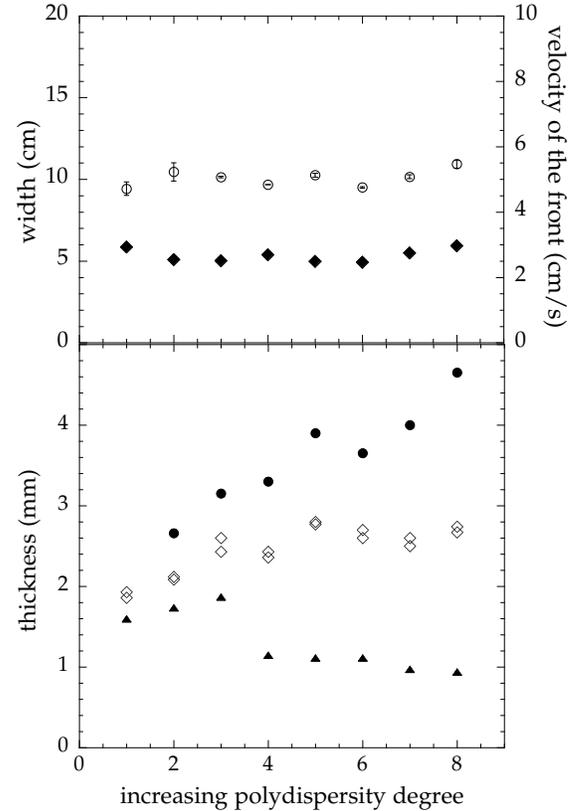}
\caption{(top) ({\Large$\circ$}) The width and ({$\blacklozenge$}) the velocity of the flow seem to be independent of the polydispersity degree. The variability of the flows is stronger than errors coming from the measurement method (error bars). The vertical axes have been chosen to compare these date to the variations induced by the range of flux we previously studied (fig. \ref{toutfctflux}). (bottom) Heights of ($\lozenge$) lev\'ees and ({\large$\bullet$}) flow increase with polydispersity degree, when ($\blacktriangle$) the thickness of the channel decreases. X-axis is arbitrary: polydispersity increases from left to right (X values correspond to batch numbers in table I) (flux: 8.15 g/s, roughness: 425-600 $\mu$m, slope: 25$^\circ$).}
\label{toutfctpoly}  
\end{figure}

Consequently, a variation of the height of the lev\'ees, or of the thickness of the channel can be explained either by a variation of the flow composition (fig. \ref{toutfctpoly}) or by a variation of the flow dynamics (fig. \ref{toutfctflux}).
The normalized relation between velocity and deposit morphology  (eq. \ref{vitdep}) has to be extended for polydisperse media before being used to quantitatively relate the dynamics and morphology in the field. 

On the other hand, the velocity increases only very slightly with the degree of polydispersity (fig. \ref{toutfctpoly}). This is surprising since it increases less (maximum 1.16) than in the case of the bidisperse media having similar fraction of fines and size range. We assume that the beads are vertically sorted in the flow: the smallest beads at the bottom, and progressively larger particles are located higher in the layer, the largest ones being at the top. In that hypothesis, the basal smoothing process looses efficiency when not applied directly to the largest particles, as for the case of a bidisperse media.  
This velocity variation has to be compared to the velocity variation induced by the flux variation. The range of fluxes and degrees of polydispersity used in our experiments are valuable to be compared because they both induce a similar variation of height. The influence of the flux (fig. \ref{toutfctflux}) and of the polydispersity degree (fig. \ref{toutfctpoly}) allows a comparison of the sensitivity of the three variables ($h$, $v$, $L$) with the two parameters (flux, polydispersity). Compared to flux-induced variation, the influence of polydispersity is negligible and  velocity could be considered as constant. In the same way, the width of the flow is constant (varying less than its fluctuations) with the polydispersity, and depends only on the flux, i.e. on the dynamics. In conclusion, width and velocity vary only slightly with polydispersity, and might be used to establish a relation which is independent of the polydispersity.

The width $L$ seems to be more promising than $h_{lev\acute{e} e}$ for making a link between the morphology of the deposit and the dynamics of the flow, which can be transposed further to the natural case. However, we will have to look for the horizontal length scale of the system, which will be the relevant parameter to normalize $L$, in the same manner that $h$ can be normalized by $h_{stop}$.

\section{Conclusions}
Good accordance between field observations and our experiments shows that some small volume pyroclastic flows behave as dense granular flows.
For these dense granular flows down rough inclines, several flow regimes have been shown to
occur depending on the inclination of the plane and on the flux imposed by the source. In each regime, the flow obeys different dynamics and is linked to a particular deposit morphology. The qualitative observation of the deposit allows the determination of the flow regime. The determination of the regime fields for a natural system could even be a step towards the establishment of an equivalence between volcanic particles flowing on a natural substratum and some laboratory granular flow conditions. In fact, one limits between regimes corresponds to the $h_{stop}(\theta)$ curve used in laboratory to normalize the equations of dense granular flow dynamics.

In the steady state frictional regime, the flow propagates with a constant height, width, and mean velocity. The associated deposit displays the lev\'ee/channel morphology often observed in the field for unconfined small pyroclastic flows. This morphology is without doubt the signature 
of the steady state frictional regime. This morphology results from static lateral zones and a partial emptying of the central part of the flow. The formation of the morphology shows that the height of lev\'ees is linked to the thickness of the flow, and it is possible to quantitatively relate the flow velocity and the morphological parameters of the deposit. 
The method is very promising for future modeling of dense pyroclastic flows because quantitative information about the flow can be extracted only from deposit measurements. A relation has been established in the case of the experimental data, for the steady state regime associated to the lev\'ee/channel morphology. The forthcoming step could be considered in two different ways. First, the relation established in the laboratory for a small range of particle sizes could be extended to the case of a highly polydisperse material. Second, width and velocity could be directly linked, because they seem to depend on dynamics of the flow, and not on polydispersity. However, we need a relevant parameter to normalize the width in order to extend the relation to the case of pyroclastic flows. 
The experimental study of the dynamics of granular flows and of the formation of the associated deposits morphology could be a way to indirectly infer data about the dynamics of small pyroclastic flows.\newline
{\bf Acknowledgments.} 
This work has been supported by the CNRS program on Natural Hazards. We thank K. Kelfoun for the Lascar global view, and A. Daerr, S. Douady, and B. Andreotti for the image analysis software.


\end{document}